\documentclass{kapproc} 
\let\footnote\savefootnote
\let\footnotetext\savefootnotetext 
\let\footnoterule\savefootnoterule 
\kluwerbib

\RequirePackage{graphicx}%
\RequirePackage{epsf}%

\newcommand{\lya}{Ly$\alpha$\ }

\newcommand{\nh}{N_{\rm HI}}

\newcommand\cdunits{{\rm cm}^{-2}}

\newcommand{\hkpc}{h^{-1}{\rm kpc}}
\newcommand{\hmpc}{h^{-1}{\rm Mpc}}

\begin{document}

\articletitle{\lya Absorber Correlations and the ``Bias" of the \lya Forest}

\author{Romeel Dav\'e$^1$, Neal Katz$^2$, \& David H. Weinberg$^3$}
\affil{$^1$ Steward Observatory, 933 N. Cherry Ave., Tucson, AZ 85721\\
$^2$ Dept. of Astronomy, Univ. of Massachusetts, Amherst, MA 01003\\
$^3$ Dept. of Astronomy, Ohio State University, Columbus, OH 43210
}

\chaptitlerunninghead{Correlations and Bias}

\begin{abstract}
\lya absorber correlations contain information about the underlying density
distribution associated with a particular class of absorbers.  As such,
they provide an opportunity to independently measure the ``bias" of
the Lyman alpha forest, i.e. the relationship between \ion{H}{1} column
density and underlying dark matter density.  In these proceedings we
use hydrodynamic simulations to investigate whether the evolution of
this bias is measurable from observable correlations.  Unfortunately, the
increasingly complex physics in the IGM at $z\la 1$ makes a direct
measurement of the bias difficult.  Nevertheless, current simulations
do make predictions for \ion{H}{1} absorber correlations that are in
broad agreement with observations at both high and low redshift, thus
reinforcing the bias evolution predictions given by these models.
\end{abstract}

Recent results indicate that hydrodynamic simulations provide an accurate
description of the local intergalactic medium as traced by weak ($\nh\la
10^{14}\cdunits$) \lya absorption lines (e.g. Dav\'e \& Tripp 2001).
A key free parameter in such models is the ``bias" of the \lya forest,
i.e. the relationship between the column density of a given absorber and
the density of the underlying dark matter associated with that absorber.
(Since pressure forces are typically small at the low densities and
temperatures in the diffuse IGM, local dark matter and baryon densities
trace each other very well.)  At high redshifts ($z\ga 2$), there is a
tight relationship between these quantities (Hui \& Gnedin 1997; Croft
et al. 1998), hence the bias is well-defined (i.e. non-stochastic, to
borrow a term from galaxy survey studies).  However, by the
present epoch, many baryons have collapsed into galaxies or been
shock heated on filaments to warm-hot temperatures (Cen \& Ostriker 1999;
Dav\'e et al. 2001), hence the relationship between \lya absorption and
the underlying mass distribution becomes more complex.  The evolution of
this bias is primarily governed by the
expansion of the universe and the strength of the photoionizing background, the
latter being the largest uncertainty in modeling \lya absorber properties
at present.  Hence specifying, or better yet measuring, the evolution
of this bias allows us construct a complete model of the \lya forest.

In these proceedings we use simulations to investigate whether it is
possible to constrain the evolution of \lya absorption bias using only
the observed correlation strength of \lya absorption.  Unfortunately,
we will see that various physical effects make this a difficult task,
at least given present observational capabilities.  Nevertheless, the
simulations make interesting predictions regarding the evolution of
\ion{H}{1} correlations from high to low redshift that are preliminarily
in agreement with observations.  Furthermore, \ion{O}{6} correlations
show interesting trends that may help to unravel their association with
\lya absorbers, as well as constrain the growth of metals in the IGM.

Our simulation results are obtained from a PTreeSPH run having $128^3$
gas and $128^3$ dark matter particles in a $22.222\hmpc$ volume with
a $5\hkpc$ softening length.  Our cosmological model is $\Lambda$CDM
($\Omega_m=0.4$) with $\Omega_b=0.02h^{-2}$, $\sigma_8=0.8$ and $h=0.65$.
At $z=2,1,0$ we extract and analyze 400 spectra along random lines of
sight through the volume.  We add Gaussian noise of $S/N=25$ to each
3~km/s pixel.  Lines are identified and fit using AutoVP (Dav\'e et
al. 1997).  Note that the total redshift path lengths at $z=2,1,0$
are $\Delta z=10.0, 5.8, 3.0$, respectively.

\section{Pixel Correlations}

\begin{figure}
\plottwo{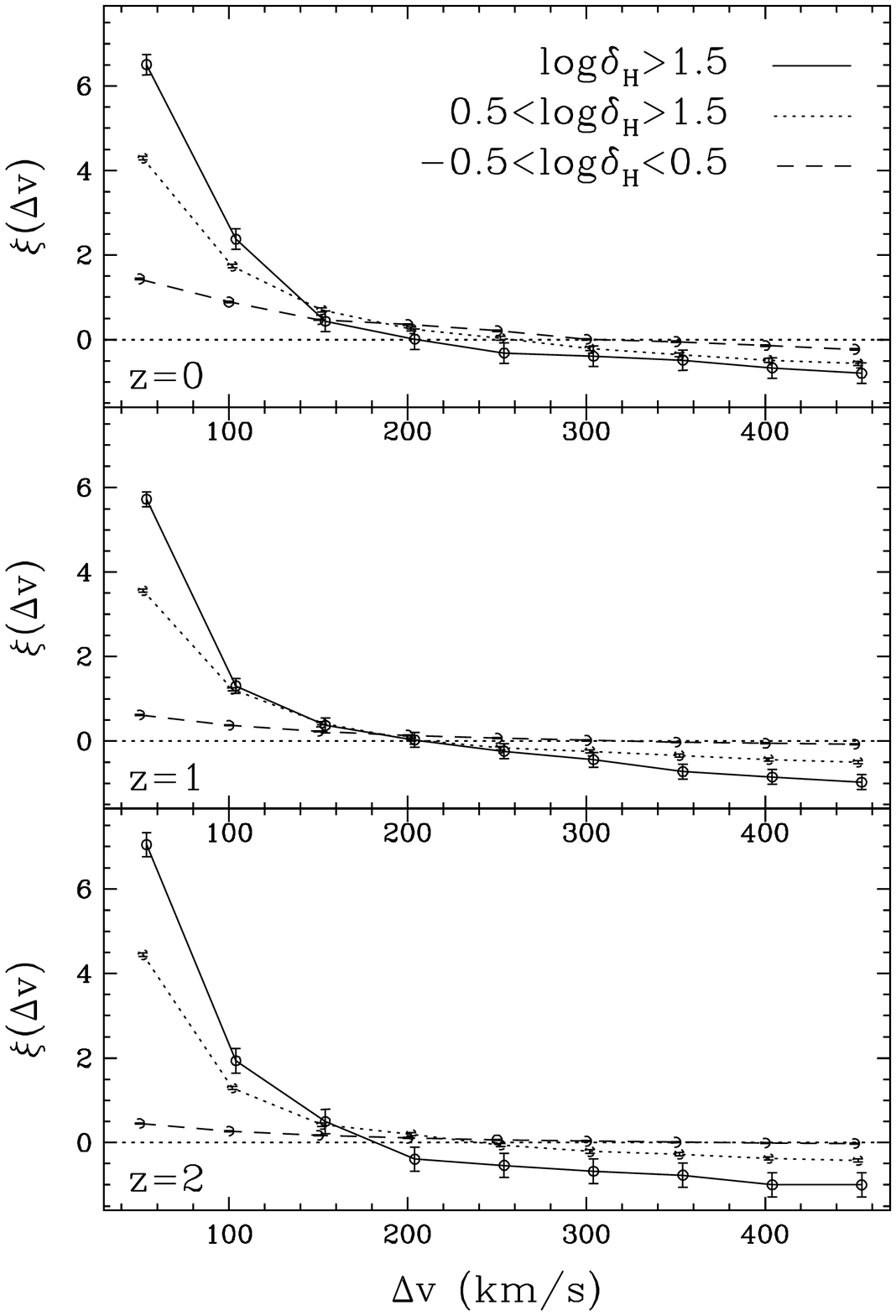}{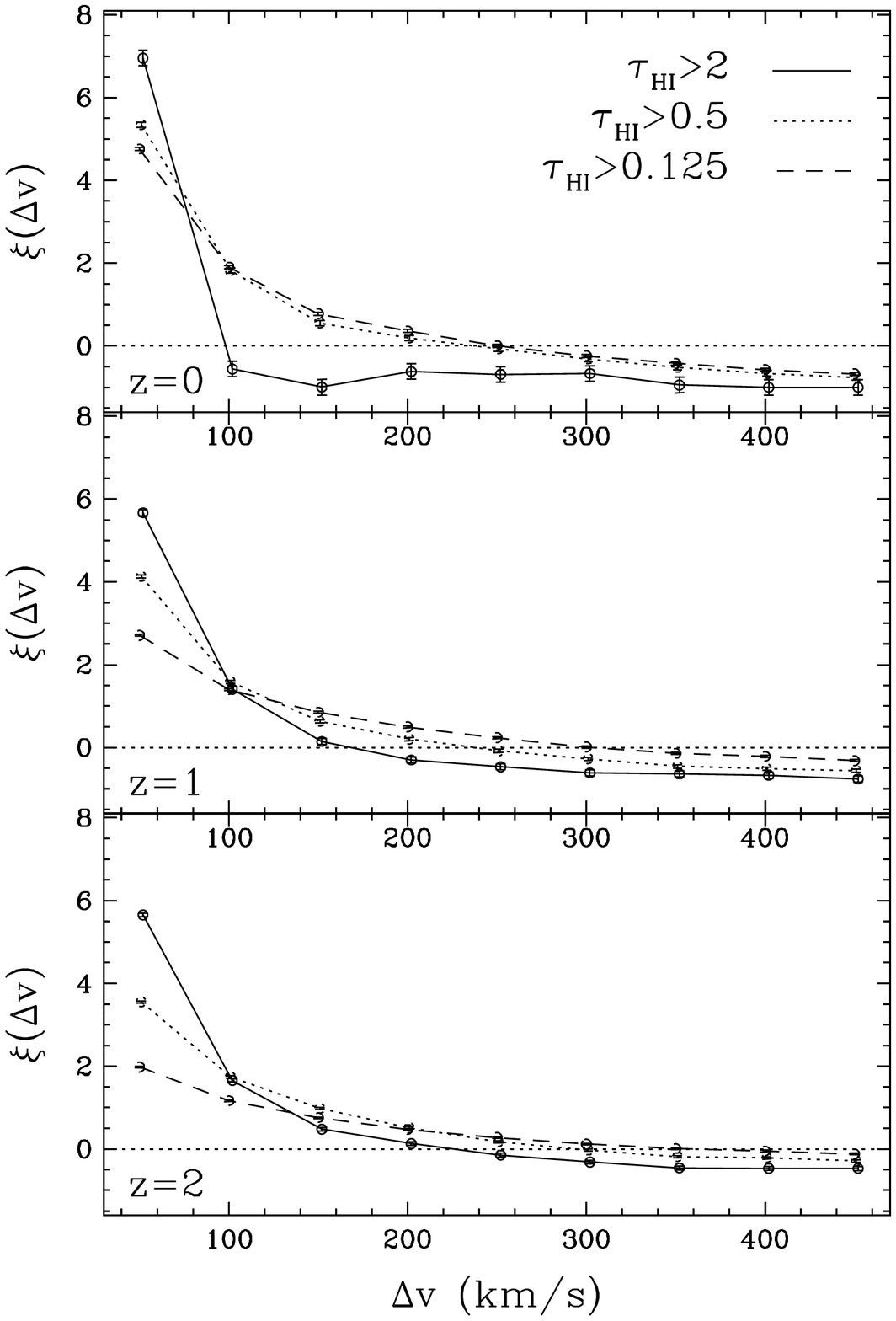}
\caption{-- {\it Left panel:} Line-of-sight density correlations at $z=0,1,2$
for pixels in the range of $-0.5<\log{\rho/\bar{\rho}}<0.5$,
$0.5<\log{\rho/\bar{\rho}}<1.5$, and $\log{\rho/\bar{\rho}}>1.5$. 
{\it Right panel:} Line-of-sight correlations of pixels with \ion{H}{1}
optical depth $\tau>0.125,0.5,2$.  
\label{fig: dencorr}}
\end{figure}

In bottom-up hierarchical structure formation models, larger overdensities
are more strongly correlated.  In Figure~\ref{fig: dencorr} (left panel)
we show how this relationship is manifested along one-dimensional
redshift-space lines of sight through our simulation.  We compute the
excess number of pairs of pixels having densities (normalized to the
cosmic mean) in the ranges typical of the \lya forest, relative to a
randomly distributed set of such pixels.  Despite smearing by peculiar
velocities, higher overdensities are still more strongly correlated,
though in velocity-space the correlation length is only $\sim 100$~km/s.
Note that the correlation {\it length} does not increase with density,
contrary to what one finds in large-scale structure studies where
e.g. the correlation length of clusters is higher than that of galaxies.
Also, it is evident that the density correlation doesn't change from
$z=2\rightarrow 0$, indicating stable clustering of \lya forest structures.

At high redshift, the density is tightly correlated with the optical
depth of \lya absorption.  Hence we would expect that \lya optical
depths would reflect correlations in the density.  The evolution of such
correlations would then, in principle, reflect the evolution of bias for a
particular optical depth, since the density correlations are not evolving.
The right panel of Figure~\ref{fig: dencorr} shows the correlations for pixels
with optical depths $\tau>0.125,0.5,2$.  According to Cen et al. (1998),
this measure most accurately reflects the underlying matter correlation,
at least at $z=3$.  Note that here we are using the noise-added spectra,
with the optical depth computed by inverting the flux, so the $\tau$
limits are actually flux limits of $F<0.88,0.61,0.14$, respectively.

Figure~\ref{fig: dencorr} (right panel) shows that the flux correlations
indeed show similar trends as density correlations at $z=2$ and 1, but
by $z=0$ the optical depth limits do not clearly delineate density cuts.
Furthermore, even from $z=2\rightarrow 1$ the correlation strength (for
$\Delta v<100$~km/s) has evolved very little, despite a comparatively
large evolution in the photoionizing background that governs the bias
(see e.g. Figure~7 in Dav\'e \& Tripp 2001).  At $z=0$, there is even a
significant {\it anti-correlation} for $\tau>2$ pixels, until $\Delta
v<100$~km/s.  Thus, disappointingly, it appears that the pixel flux
(or optical depth) correlations do not straightforwardly trace the 
evolution of the \lya forest bias.

\section{Line Correlations}

\begin{figure}
\plottwo{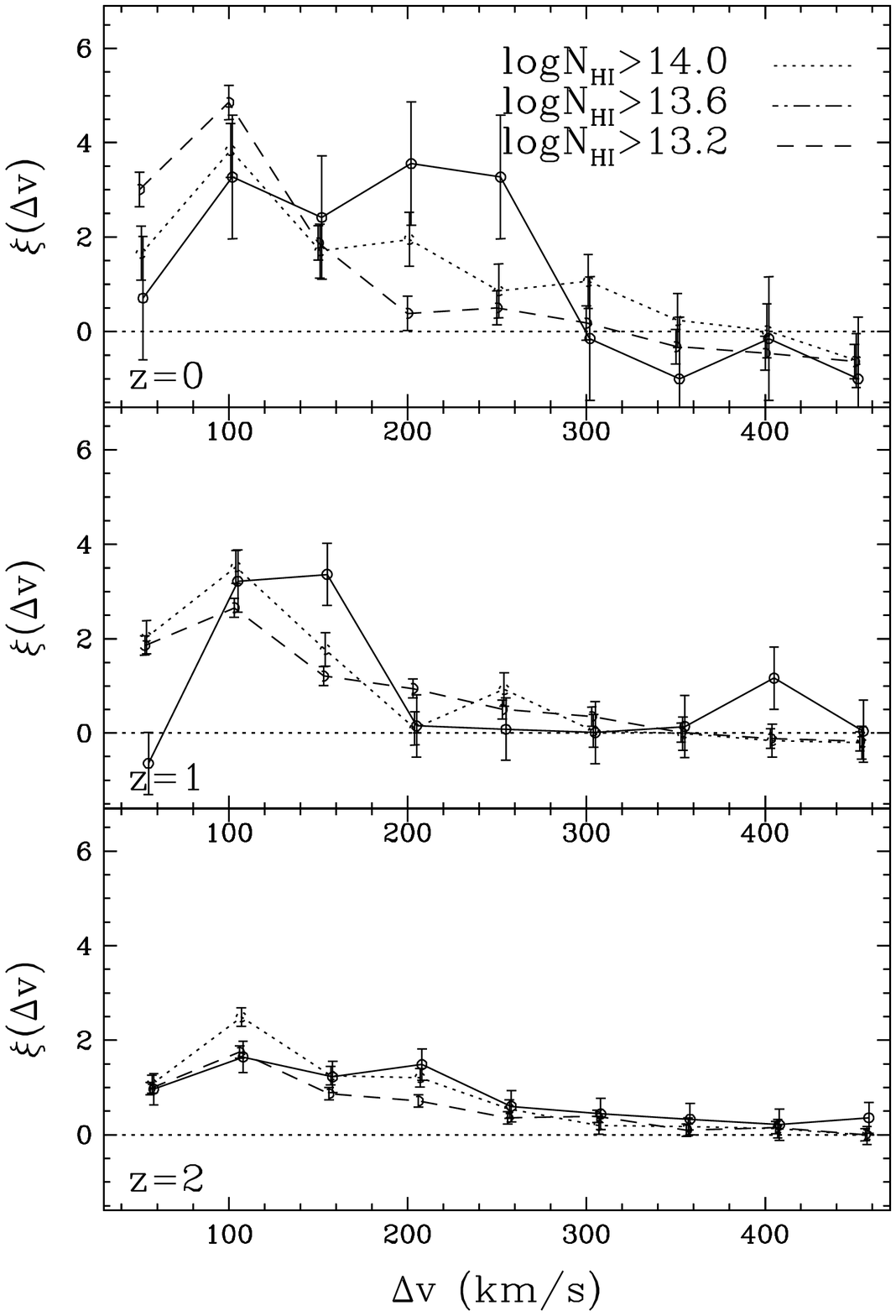}{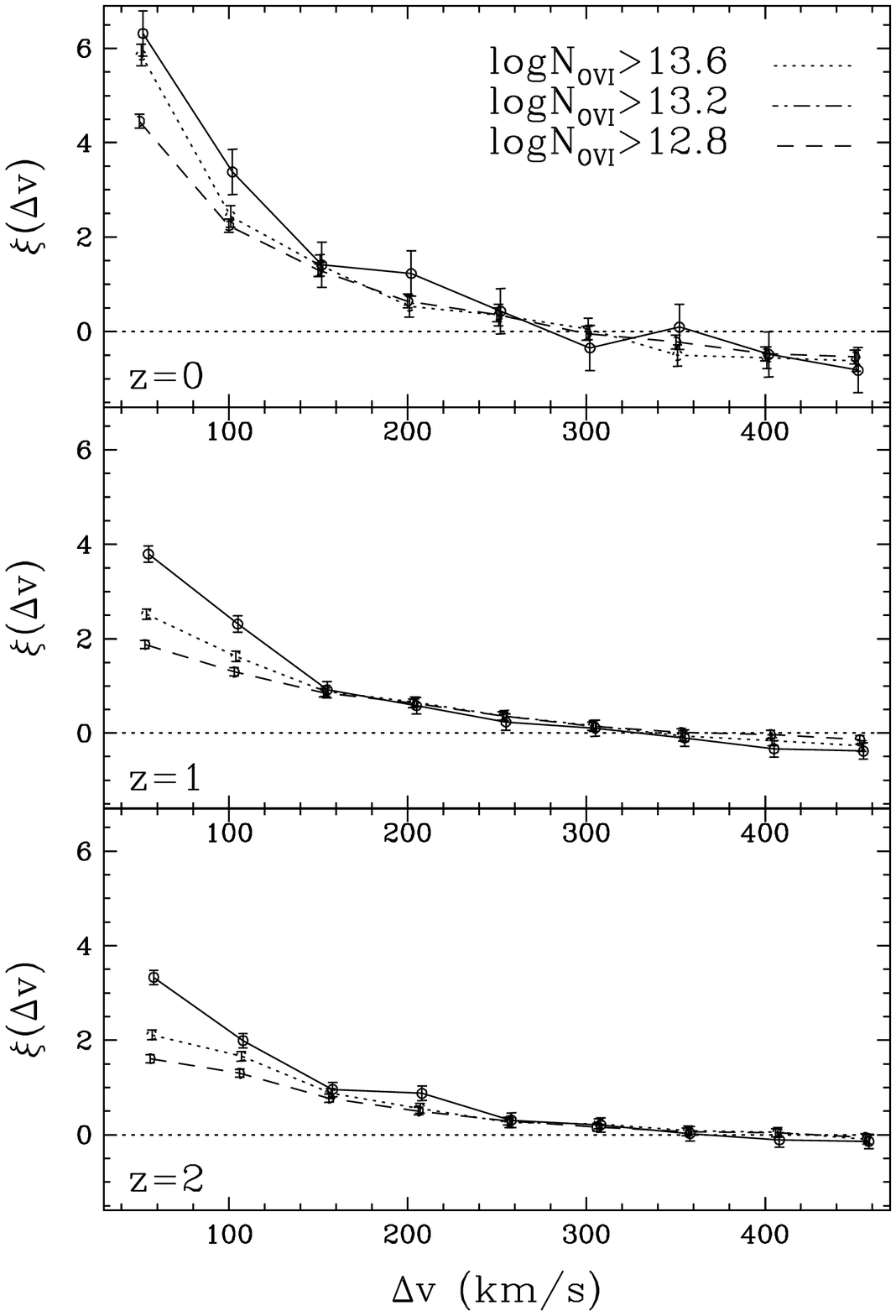}
\caption{-- {\it Left panel:} Line correlation function for \lya 
at $z=0,1,2$, with limits of $\log{\nh}>13.2,13.6,14$. 
{\it Right panel:} Same for 
\ion{O}{6} absorbers, with limits of $\log{N_{\rm OVI}}>12.8,13.2,13.6$.
Note: Error bars shown are statistical,
and do not include cosmic variance which is likely to dominate the
error budget.
\label{fig: corr}}
\end{figure}

The canonical method for studying \lya absorbers is by profile-fitting
individual features.  At high redshifts, studies indicate that
\lya lines are uncorrelated for $\nh\la 10^{14}\cdunits$.  In these
proceedings, low-redshift STIS spectra of Tripp et al. and Williger
et al. show significant excess correlations of \ion{H}{1} lines
over a random distribution out to $250-300$~km/s, for absorbers with
$\nh\ga 10^{13.6}\cdunits$.  The implication is that the correlation
strength at this column density has increased with time, which is
(qualitatively) expected from simulations since a given column density
absorber is associated with higher overdensities at lower redshifts
(Dav\'e et al. 1999).  Conversely, a poster by Heap et al. indicates no
significant absorber correlations in a sightline towards 3C273 ($z_{\rm
em}=0.156$), where the typical absorber has a lower column density
($\nh\sim 10^{13}\cdunits$).  These results lend broad support to the
bias evolution model given by simulations, but at present are insufficient
to precisely constrain the bias evolution from high to low redshifts.

Figure~\ref{fig: corr} (left panel) shows our line correlation function
for various column density limits, at $z=0,1,2$.  The line correlations at
a given $\nh$ grow with time, in agreement with observations.  At $z=2$,
virtually no correlations are seen in any column density range, while
at $z=0$ significant correlations are seen out to $\Delta v\approx
300$~km/s, for stronger lines.  These trends are also in broad agreement
with observations.  Note also the interesting drop in correlation strength
at $\Delta v\la 50$~km/s; this is likely sensitive to line deblending
algorithms, but may be an interesting regime for comparing simulations
to observations if identical profile fitting routines are used.

Still, there seems to be no direct relationship between correlations
computed from density cuts (cf. Figure~\ref{fig: dencorr}) and from
lines with column density cuts, meaning that while current simulations
broadly reproduce observed line correlations, such correlations are also
not simply related to the bias of the \lya forest.

\section{\ion{O}{6} Line Correlations}

Recent observations suggest a significant number of \ion{O}{6} lines
present in the local universe.  Such lines may arise in collisionally
ionized ``warm-hot" gas or in very low density photoionized gas.
Simulations roughly match the observed number density per unit redshift
of lines by assuming [O/H]$\sim -1$ and a quasar-dominated flux (e.g. Chen
et al. 2002), with stronger absorbers tending
to be collisionally ionized and the weaker absorbers photoionized
(Cen et al. 2001, Fang \& Bryan 2001).  The correlation of \ion{O}{6}
lines can, in principle, be used as a diagnostic to determine their
origin, based on a comparison of their clustering strength with that
of \ion{H}{1} absorbers.  A complication is that the metallicity and
far-UV ionization conditions of the IGM are poorly determined.

We compute \ion{O}{6} line correlations from our simulation by assuming
a Haardt \& Madau (1996) ionizing background and a spatially-uniform
metallicity that grows with time: [O/H]$=-2$ at $z=2$, [O/H]$=-1.5$ at
$z=1$, and [O/H]$=-1$ at $z=0$.  These assumptions are reasonable but
fairly arbitrary, as observational constraints are poor (see Prochaska, these
proceedings, for a review).  We also extract spectra with only \ion{O}{6}
absorption, neglecting the real-world complication of blending with
\ion{H}{1}.  

Figure~\ref{fig: corr} (right panel) shows the resulting \ion{O}{6}
correlation function based on these assumptions.  It is clear that
stronger \ion{O}{6} lines are more strongly correlated, particularly
at $z=1$ and 2.  This indicates that many of these \ion{O}{6} lines
are photoionized, because collisionally ionized absorbers should have
their column density virtually uncorrelated with density.
At $z=0$, this is not so clear, indicating more lines at these column
densities may be collisionally ionized.  Furthermore, the correlations
of these lines are fairly strong out to hundreds of km/s, indicating that
\ion{O}{6} absorption is mainly occuring in filaments.  Finally, with the stated
assumptions, the \ion{O}{6} line correlation strength does not increase
significantly with redshift in comparison with that of \ion{H}{1}.

At present there are insufficient numbers of \ion{O}{6} absorbers observed
to test these simulation predictions.  Upcoming observations with COS may
alleviate this situation, though issues of blending with \ion{H}{1} and
uncertainties in ionization conditions will make interpretation difficult.
In principle, a similar analysis could be applied to \ion{C}{4} absorbers;
while not shown for lack of space, \ion{C}{4} also shows very little
evolution in correlation strength (assuming an increasing metallicity
with time), and even stronger correlations than \ion{O}{6} at all redshifts.

\section{Conclusions}

We have investigated various line-of-sight autocorrelation measures
for weak \ion{H}{1} and \ion{O}{6} absorption in the IGM.  Correlations
can in principle associate a given optical depth or column density
with underlying physical densities within a hierarchical framework,
thereby constraining the ``bias" of the \lya forest.  Unfortunately, the
increasingly complex physics associated with the low-redshift IGM make
this untenable at present.  Nevertheless, predictions of line correlations
from hydrodynamic simulations broadly agree with observations, showing a
growing correlation strength with time at fixed a fixed column density,
and significant correlations out to $\sim 300$~km/s for strong
lines at $z\sim 0$.  More careful comparisons and improved observations
will be needed to quantitatively assess any discrepancies, but
for now it appears that the bias evolution model forwarded by structure
formation scenarios for the \lya forest is in agreement with observations.



\begin{chapthebibliography}{1}
\noindent Cen, R., Phelps, S., Miralda-Escud\'e, J., \& Ostriker, J. P. 1998, ApJ, 496, 577 \\
\noindent Cen, R. \& Ostriker, J. P. 1999, ApJ 514, 1 \\
\noindent Cen, R. Tripp, T. M., Ostriker, J. P., \& Jenkins, E. B. 2001, ApJ, 559, L5 \\
\noindent Chen, X., Weinberg, D. H., Katz, N., \& Dav\'e, R. 2002, ApJ, accepted \\
\noindent Croft, R. A. C., Weinberg, D. H., Katz, N., \& Hernquist, L. 1998, ApJ, 495, 44 \\
\noindent Dav\'e, R., Hernquist, L., Weinberg, D. H., \& Katz, N. 1997, ApJ, 477, 21 \\
\noindent Dav\'e, R., Hernquist, L., Katz, N., \& Weinberg, D. H. 1999, ApJ, 511, 521 \\
\noindent Dav\'e, R. et al. 2001, ApJ, 552, 473 \\
\noindent Dav\'e, R. \& Tripp, T. M. 2001, ApJ, 553, 528 \\
\noindent Fang, T. \& Bryan, G. L. 2001, ApJ, 561, L31 \\
\noindent Hui, L. \& Gnedin, N. 1997, MNRAS, 292, 27 \\
\end{chapthebibliography}

\end{document}